%% file: ISIT2009.tex
\documentclass[conference]{IEEEtran}\IEEEoverridecommandlockouts

\usepackage{graphicx}
\usepackage{amsmath,amssymb} \interdisplaylinepenalty=2500
\usepackage{cite}
\usepackage{stmaryrd}
\usepackage{verbatim}
\usepackage{multirow}

\newtheorem{definition}{Definition}
\newtheorem{theorem}{Theorem}

\newtheorem{lemma}[theorem]{Lemma}

\def\qed{\endIEEEproof}

\DeclareMathOperator{\rank}{\sf rank\hspace{0.1em}}

\DeclareMathOperator{\Tr}{\sf Tr}

\newcommand{\Fq}{\mathbb{F}_q}
\newcommand{\Fqm}{\mathbb{F}_{q^m}}

\newcommand{\mat}[1]{\begin{bmatrix} #1 \end{bmatrix}}

\newcommand{\bbrack}[1]{\left\llbracket #1 \right\rrbracket}
\newcommand{\ul}{\underline}

\newcommand{\dr}{d_{\textrm{R}}}

\input{mymacros}

\title{Fast Encoding and Decoding of Gabidulin Codes}

\author{
\IEEEauthorblockN{Danilo Silva and Frank R. Kschischang}
\IEEEauthorblockA{Department of Electrical and Computer Engineering, University of Toronto \\
Toronto, Ontario M5S 3G4, Canada, {\{danilo, frank\}@comm.utoronto.ca}\vspace{-0.9ex}
}
\thanks{This work was supported by CAPES Foundation, Brazil.}
}

\begin{document}
\maketitle
\thispagestyle{empty}

\input{section0}
\input{section1}

\input{section2}

\input{section3}

\input{section4}

\input{section5}

\input{section7}

\bibliographystyle{IEEEtran}
\bibliography{IEEEabrv,networkcoding,codingtheory,rankmetric,silva,books,finitefields}

\end{document}

%% file: mymacros.tex
\newcommand{\calA}{\mathcal{A}}
\newcommand{\calB}{\mathcal{B}}
\newcommand{\calC}{\mathcal{C}}

\newcommand{\calS}{\mathcal{S}}

%% file: section0.tex
\begin{abstract}
Gabidulin codes are the rank-metric analogs of Reed-Solomon codes and have a major role in practical error control for network coding. This paper presents new encoding and decoding algorithms for Gabidulin codes based on low-complexity normal bases. In addition, a new decoding algorithm is proposed based on a transform-domain approach. Together, these represent the fastest known algorithms for encoding and decoding Gabidulin codes.

\end{abstract}

%% file: section1.tex
\section{Introduction}
\label{sec:introduction}

Gabidulin codes \cite{Gabidulin1985} are optimal codes for the rank metric that are closely related to Reed-Solomon codes. These codes have attracted significant attention recently as they can provide a near-optimal solution to the error control problem in network coding \cite{Kotter.Kschischang2008,Silva++2008}.

So far, two methods have been proposed for decoding Gabidulin codes: a ``standard'' method based on the Berlekamp-Massey algorithm (or the extended Euclidean algorithm) \cite{Gabidulin1985,Richter.Plass2004:BerlekampMassey}, and a method based on a Welch-Berlekamp key equation \cite{Loidreau2005}. The two methods are most efficient for high-rate and low-rate codes, respectively \cite{Gadouleau.Yan2008a:Complexity}.

In this paper, we improve the computational complexity of the standard (time-domain) algorithm by the use of optimal or low-complexity normal bases. With this modification, the two most demanding steps (computing the syndromes and finding the root space of the error span polynomial) become quite easy to perform. In addition, we propose a transform-domain approach to the decoding, based on a novel definition of a Fourier-like transform for vectors in a finite extension field. The transform-domain approach is shown to be more suitable for low-rate codes, while the time-domain method is most suitable for high-rate codes. Drawing on the insights above, we also propose two new encoding algorithms, which improve on the complexity for either systematic high-rate codes or nonsystematic codes.

The transform approach has the additional benefit of providing new, simpler proofs of the key equations, further strengthening the connections with classical coding theory.

The remainder of the paper is organized as follows. In Section~\ref{sec:preliminaries}, we review rank-metric codes and some known results about normal bases.
Sections~\ref{sec:decoding},~\ref{sec:transform-domain}~and~\ref{sec:encoding} present, respectively, our time-domain decoding algorithm, our transform-domain decoding algorithm, and our two encoding algorithms. Finally, Section~\ref{sec:conclusions} presents our conclusions. Most proofs have been omitted due to lack of space. 

%% file: section2.tex
\section{Preliminaries}
\label{sec:preliminaries}


%

\subsection{Linear Algebra Notations}
\label{ssec:linear-algebra}

In this paper, all bases, vectors and matrices are indexed starting from 0.
Let $V$ and $W$ be finite-dimensional vector spaces over a field $F$ with ordered bases $\calA=\{\alpha_0,\ldots,\alpha_{n-1}\}$ and $\calB=\{\beta_0,\ldots,\beta_{m-1}\}$, respectively. For $v \in V$, we denote by $\mat{v}_\calA$ the coordinate vector of $v$ relative to $\calA$; that is, $\mat{v}_\calA = \mat{v_0 & \cdots & v_{n-1}}$, where $v_0,\ldots,v_{n-1}$ are the unique elements in $F$ such that $v = \sum_{i=0}^{n-1} v_i \alpha_i$. Let $T$ be a linear transformation from $V$ to $W$. We denote by $\mat{T}_{\calA}^{\calB}$ the matrix representation of $T$ in the bases $\calA$ and $\calB$; that is, $\mat{T}_{\calA}^{\calB}$ is the unique $n \times m$ matrix over $F$ such that
  $T(\alpha_i) = \sum_{j=0}^{m-1} \left(\mat{T}_{\calA}^{\calB}\right)_{ij} \beta_j, \quad i=0,\ldots,n-1.$
With these notations, we have $\mat{T(v)}_\calB = \mat{v}_\calA \mat{T}_{\calA}^{\calB}$. Let $U$ be a finite-dimensional vector space over $F$ with ordered basis $\Theta = \{\theta_0,\ldots,\theta_{k-1}\}$, and let $S$ be a linear transformation from $W$ to $U$. Recall that \cite{Friedberg++}
\begin{equation}\label{eq:matrix-multiplication-linear-maps}
  \mat{TS}_{\calA}^{\Theta} = \mat{T}_{\calA}^{\calB} \mat{S}_{\calB}^{\Theta}.
\end{equation}

\subsection{Rank-Metric Codes}
\label{ssec:rank-metric}

Let $q$ be a power of a prime and let $\Fq$ denote the finite field with $q$ elements. Let $\Fq^{n \times m}$ denote the set of all $n \times m$ matrices over $\Fq$, and set $\Fq^n = \Fq^{n \times 1}$. Let $\Fqm$ be an extension field of $\Fq$. Recall that every extension field can be regarded as a vector space over the base field. Let $\calA = \{\alpha_0,\ldots,\alpha_{m-1}\}$ be a basis for $\Fqm$ over $\Fq$.
Since $\Fqm$ is also a field, we may consider a vector $v \in \Fqm^n$. Whenever $v \in \Fqm^n$, we denote by $v_i$ the $i$th entry of $v$; that is, $v = \mat{v_0 & \cdots & v_{n-1}}^T$. It is natural to extend the map $\mat{\cdot}_\calA$ to a bijection from $\Fqm^n$ to $\Fq^{n \times m}$, where the $i$th row of $\mat{v}_\calA$ is given by $\mat{v_i}_\calA$. When the basis $\calA$ is fixed, we will use the simplified notation $\underline{a} \triangleq \mat{a}_\calA$ for $a \in \Fqm$ and $\underline{v} \triangleq \mat{v}_\calA$ for $v \in \Fqm^n$.

Define the rank of a vector $v \in \Fqm^n$, denoted by $\rank(v)$, to be the rank of the associated matrix $\underline{v}$, that is, $\rank(v) \triangleq \rank(\underline{v})$.
Similarly, define the \emph{rank distance} between $u,v \in \Fqm^n$ to be $\dr(u,v) \triangleq \rank(\underline{v} - \underline{u})$. It is well-known that the rank distance is indeed a metric on $\Fqm^n$ \cite{Gabidulin1985}.

A rank-metric code $\calC \subseteq \Fqm^n$ is a block code of length $n$ over $\Fqm$ that is well-suited to the rank metric. We use $\dr(\calC)$ to denote the minimum rank distance of $\calC$.


For $m \geq n$, an important class of rank-metric codes was proposed by Gabidulin \cite{Gabidulin1985}. Let $[i]$ denote $q^i$. A \emph{Gabidulin code} is a linear $(n,k)$ block code over $\Fqm$ defined by the parity-check matrix $H = \mat{h_{j}^{[i]}}$, $0 \leq i \leq n-k-1$, $0 \leq j \leq n-1$,
where the elements $h_0,\ldots,h_{n-1} \in \Fqm$ are linearly independent
over $\Fq$. It can be shown that the minimum rank distance of a Gabidulin code is $d = n-k+1$, so the code satisfies the Singleton bound for the rank metric \cite{Gabidulin1985}.

\subsection{Linearized Polynomials}
\label{ssec:linearized-polynomials}


A \emph{linearized polynomial} or \emph{$q$-polynomial} over $\Fqm$ \cite{Lidl.Niederreiter} is a polynomial of the form
 $f(x) = \sum_{i=0}^{n} f_i x^{[i]}$,
where $f_i \in \Fqm$. If $f_n \neq 0$, we call $n$ the \emph{$q$-degree} of $f(x)$. It is easy to see that evaluation of a linearized polynomial is
 an $\Fq$-linear transformation from $\Fqm$ to itself. In particular, the set of roots in $\Fqm$ of a linearized polynomial is the kernel of the associated map (and therefore a subspace of $\Fqm$).

It is well-known that the set of linearized polynomials over $\Fqm$ forms an $\Fq$-algebra under addition and composition (evaluation). The latter operation is usually called \emph{symbolic multiplication} in this context and denoted by $f(x) \otimes g(x) = f(g(x))$. Note that if $n$ and $k$ are the $q$-degrees of $f(x)$ and $g(x)$, respectively, then the $q$-degree of $f(x) \otimes g(x)$ is equal to $n + k$.

Let $\calS \subseteq \Fqm$. The $q$-polynomial $M_\calS(x) = \sum_{i=0}^{t} M_i x^{[i]}$ with $M_0 = 1$ and least $q$-degree $t$ whose root space contains $\calS$ is unique and is called the \emph{minimal $q$-polynomial} of $\calS$. The $q$-degree of $M_\calS(x)$ is precisely equal to the dimension of the space spanned by $\calS$, and is also equal to the nullity of $M_\calS(x)$ as a linear map.

\subsection{Normal Bases}
\label{ssec:normal-bases}

If a basis $\calA$ for $\Fqm$ over $\Fq$ is of the form $\calA = \{\alpha^{[0]},\alpha^{[1]},\ldots,\alpha^{[m-1]}\}$, then $\calA$ is called a \emph{normal basis} and $\alpha$ is called a \emph{normal element} \cite{Lidl.Niederreiter}. Assume that $\calA$ is fixed.
Let $\bbrack{\alpha}$ denote the column vector $\mat{\alpha^{[0]} & \cdots & \alpha^{[m-1]}}^T$. Then any element $a \in \Fqm$ can be written as $a = \underline{a} \bbrack{\alpha}$.

For a vector $\ul{a} = \mat{a_0,\ldots,a_{m-1}} \in \Fq^{1 \times m}$, let $\ul{a}^{\leftarrow i}$ denote a cyclic shift to the left by $i$ positions, that is, $\ul{a}^{\leftarrow i} = \mat{a_i,\ldots,a_{m-1},a_0,\ldots,a_{i-1}}$. Similarly, let $\ul{a}^{\rightarrow i} = \ul{a}^{\leftarrow m-i}$. In this notation, we have $a^{[i]} = \ul{a}^{\rightarrow i} \bbrack{\alpha}$, or $\ul{a^{[i]}} = \ul{a}^{\rightarrow i}$. That is, $q$-exponentiation in a normal basis is simply a cyclic shift.

Multiplications in normal bases are usually performed in the following way. Let $T = [T_{ij}] \in \Fq^{n \times m}$ be a matrix such that $\alpha \alpha^{[i]} = \sum_{j=0}^{m-1} T_{ij} \alpha^{[j]}$, $i=0,\ldots,m-1$. The matrix $T$ is called the \emph{multiplication table} of the normal basis. The number of nonzero entries in $T$ is denoted by $C(T)$ and is called the \emph{complexity} of the normal basis \cite{Gao1993:Thesis}. Note that $\alpha \bbrack{\alpha} = T \bbrack{\alpha}$. It can be shown that, if $a,b \in \Fqm$, then
  $\ul{ab} = \sum_{i=0}^{m-1} b_i \left( \ul{a}^{\leftarrow i} T\right)^{\rightarrow i}.$

Thus, a general multiplication in a normal basis requires $m C(T) + m^2$ multiplications and $m C(T) - 1$ additions in $\Fq$. Clearly, this is only efficient if $T$ is sparse; otherwise, it is more advantageous to convert back and forth to a polynomial basis to perform multiplication.

It is a well-known result that the complexity of a normal basis is lower bounded by $2m-1$. Bases that achieve this complexity are called \emph{optimal}.
%
More generally, low-complexity (but not necessarily optimal) normal bases can be constructed using \emph{Gauss periods}, as described in detail in \cite{Gao1993:Thesis}. For $q=2^s$, such a construction is possible if and only if $m$ satisfies $\gcd(m,s)=1$ and $8 \nmid m$ \cite{Gao++2000:AlgorithmsExponentiation}.
As an example, for $q=256$, this condition is satisfied for any odd $m$. Among the odd $m \leq 100$, the normal bases that result are in fact optimal when $m=3$, 5, 9, 11, 23, 29, 33, 35, 39, 41, 51, 53, 65, 69, 81, 83, 89, 95, 99. 
For $q=2^s$ and odd $m$, all of the normal bases constructed by Gauss periods are self-dual \cite{Gao++2000:AlgorithmsExponentiation}.

An interesting fact about a normal basis constructed via Gauss periods is that its multiplication table $T$ lies entirely in the prime field $\mathbb{F}_p$, where $p$ is the characteristic of $q$.
This in turn
implies that the minimal polynomial of $\alpha$ is in $\mathbb{F}_p[x]$ and the conversion matrices from/to the standard basis $\{\alpha^0,\alpha^1,\ldots,\alpha^{m-1}\}$ are also in $\mathbb{F}_p^{m \times m}$.

In this paper, we are mostly interested in the case $p=2$. In this case, multiplication by $T$ can be done simply by using XORs.
In Table~\ref{tab:operations-ext-field},
\begin{table}
  \centering
  \caption{Complexity of Operations in $\Fqm$ (upper bound)}\label{tab:operations-ext-field}
  \begin{tabular}{|l|c|c|c|}
    \hline
      \multirow{2}{*}{
      Operations in $\Fqm$
      } & \multicolumn{3}{|c|}{Number of operations in $\Fq$} \\
      \cline{2-4}
      & Multiplications & Additions & Inversions \\
      \hline
    Multiplication & $m^2$ & $m(C(T)-1)$ & -- \\
    Addition & -- & $m$ & -- \\
    Inversion & $\frac{5}{2}m^2 + O(m)$ & $4m^2 + O(m)$ & $m+2$ \\
    \hline
  \end{tabular}
  \vspace{-1ex}
\end{table}
we give the complexity of each operation in $\Fqm$ assuming that $p=2$. We also assume that $q$-exponentiations are free. Inversion is performed using the extended Euclidean algorithm on a standard basis. Details of these calculations can be found in \cite{Gathen.Gerhard}. 

%% file: section3.tex
\section{Fast Decoding of Gabidulin Codes}
\label{sec:decoding}

In this section, we assume a fixed basis $\calA$ for $\Fqm$ over $\Fq$.

\subsection{Standard Decoding Algorithm}
\label{ssec:decoding-gabidulin}

We review below the standard decoding algorithm for Gabidulin codes. This is the fastest decoding algorithm to date, except for low rates (see \cite{Gadouleau.Yan2008a:Complexity}). For details we refer the reader to \cite{Gabidulin1985,Richter.Plass2004:BerlekampMassey,Silva++2008,Gadouleau.Yan2008a:Complexity} and references therein.

Let $\calC \subseteq \Fqm^n$ be a Gabidulin code with $\dr(\calC) = d$ defined by the parity-check matrix $H = \mat{h_{j}^{[i]}}$. Let $c \in \calC$ be the transmitted word, let $e \in \Fqm^n$ be an error word of rank $\tau \leq (d-1)/2$, and let $r = c + e$ be the received word. The decoding problem is to find the unique $e$ such that $r - e \in \calC$.

Since $\rank e = \tau$, we can rewrite $e$ as
\begin{equation}\label{eq:error-expansion}
  e = LV = \mat{L_1 & \cdots & L_\tau} \mat{V_1 \\ \vdots \\ V_\tau} = \sum_{j=1}^\tau L_j V_j
\end{equation}
where $L_1,\ldots,L_\tau \in \Fq^n$ are called the \emph{error locations} and $V_1,\ldots,V_\tau \in \Fqm$ are called the \emph{error values}. (Note that this expansion is not unique.) Define the \emph{error locators}
  $X_j = L_j^T h$,
where $h = \mat{h_0,\ldots,h_{n-1}}^T \in \Fqm^n$. Define also the \emph{syndromes}
  $S_\ell = \sum_{i=0}^{n-1} h_i^{[\ell]} r_i, \quad \ell=0,\ldots,d-2.$
The error locators and error values must satisfy the \emph{syndrome equation}
\begin{equation}\label{eq:syndrome-equation}
  S_\ell = \sum_{j=1}^\tau X_j^{[\ell]} V_j ,\quad \ell = 0,\ldots,d-2
\end{equation}
or, equivalently, 
\begin{equation}\label{eq:syndrome-equation-reversed}
 \tilde{S}_\ell \triangleq S_{d-2-\ell}^{[\ell-d+2]} = \sum_{j=1}^\tau V_j^{[\ell-d+2]} X_j ,\quad \ell = 0,\ldots,d-2.
\end{equation}

The solution of the syndrome equation can be facilitated by the use of linearized polynomials. Due to the similarity between error locators and error values, there are two equivalent approaches to the problem. Define the the \emph{error span polynomial} (ESP) $\Gamma(x)$ as the minimal $q$-polynomial of $V_1,\ldots,V_\tau$ and the \emph{error locator polynomial} (ELP) $\Lambda(x)$ as the minimal $q$-polynomial of $X_1,\ldots,X_\tau$. Then, either of the following \emph{key equations} may be used:
\begin{align}
\sum_{i=0}^{\tau} \Gamma_{i} S_{\ell-i}^{[i]} &= 0, \quad \ell = \tau,\ldots, d-2 \label{eq:key-equation-ESP} \\
\sum_{i=0}^{\tau} \Lambda_{i} \tilde{S}_{\ell-i}^{[i]} &= 0, \quad \ell = \tau,\ldots, d-2. \label{eq:key-equation-ELP}
\end{align}
These key equations can be solved, for instance, with the modified Berlekamp-Massey (BM) algorithm \cite{Richter.Plass2004:BerlekampMassey}.

Assume the ESP is used. To find a basis $V_1,\ldots,V_\tau$ for the root space of $\Gamma(x)$, we can first compute the matrix $\ul{\gamma} = \mat{\Gamma(x)}_{\calA}^\calA$ representing $\Gamma(x)$ as a linear map, and then use Gaussian elimination to find a basis for the left null space of $\ul{\gamma}$. To find the error locators, we can use Gabidulin's algorithm, which is an algorithm to solve a system of the form (\ref{eq:syndrome-equation}).

Alternatively, if the ELP is used, we can use exactly the same procedure to find a basis $X_1,\ldots,X_\tau$ for the root space of $\Lambda(x)$, followed by Gabidulin's algorithm to solve (\ref{eq:syndrome-equation-reversed}) and find the error values.

After $X_1,\ldots,X_\tau$ and $V_1,\ldots,V_\tau$ are found, the error locations can be computed by
  $L_j^T = \ul{X_j} (\ul{h})^{\dagger}$,
where $(\ul{h})^{\dag}$ is a right-inverse of $\ul{h}$. Note that $(\ul{h})^{\dag}$ can be precomputed. Finally, the error word is computed from (\ref{eq:error-expansion}).

A summary of the algorithm and breakdown of complexity is given below. The algorithm consists of six steps:

\begin{enumerate}
  \item \label{step:syndromes} \emph{Compute the syndromes:} $(d-1)n$ multiplications and $(d-1)(n-1)$ additions in $\Fqm$.
  \item \label{step:ESP-ELP} \emph{Compute the ESP/ELP:} $(d-1)(d-2)$ multiplications, $\frac{1}{2}(d-1)(d-2)$ additions and $\frac{1}{2}(d-1)$ inversions in~$\Fqm$.
  \item \label{step:basis-root-space} \emph{Find a basis for the root space of the ESP/ELP:}
  \begin{enumerate}
    \item \label{step:matrix-linear-map} \emph{Compute the matrix of the linear map:} $\tau m$ multiplications and additions in $\Fqm$;
    \item \label{step:left-null-space} \emph{Compute the left null space of this matrix:} $\frac{1}{2}(m-\tau)(m+\tau-1)m$ multiplications and $\frac{1}{2}(m-\tau)(m+\tau-1)(m-1)$ additions in $\Fq$.
  \end{enumerate}
  \item \label{step:Gabidulin-algorithm} \emph{Find the error locators/error values:} 
      $\frac{3}{2}\tau^2+\frac{1}{2}\tau -1$
      multiplications, $\frac{3}{2}\tau(\tau-1)$ additions and $\tau$ inversions~in~$\Fqm$.
  \item \label{step:error-locations} \emph{Compute the error locations:} $\tau n m$ multiplications and $\tau n (m-1)$ additions in $\Fq$.
  \item \label{step:error-word} \emph{Compute the error word:} $\tau n m$ multiplications and $(\tau-1)nm$ additions in $\Fq$.
\end{enumerate}

Except for Step 3, the details of these calculations can be found in \cite{Gadouleau.Yan2008a:Complexity}. It can be seen that the complexity is dominated by steps \ref{step:syndromes} and \ref{step:matrix-linear-map}, each requiring $O(dm^3)$ operations in $\Fq$.

\subsection{Fast Decoding Using Low-Complexity Normal Bases}
\label{ssec:fast-decoding-normal-bases}

We now assume that $\calA = \{\alpha^{[i]}\}$ is a low-complexity normal basis with multiplication table $T$, and that $\Fq$ has characteristic $2$.
The essence of our approach lies in the following expression:

\begin{equation}\nonumber
  \ul{a \alpha^{[i]}} = \left( \ul{a}^{\leftarrow i} T\right)^{\rightarrow i}, \quad \forall i=0,\ldots,m-1.
\end{equation}

In other words, multiplying an element of $\Fqm$ by a $q$-power of $\alpha$ costs only $C(T)-m$ \emph{additions} in $\Fq$ (recall that $T$ lies in $\mathbb{F}_2$), rather than $O(m^2)$ operations in $\Fq$ as in a general multiplication.
Below, we exploit this fact in order to significantly reduce the complexity of decoding Gabidulin codes.

Consider, as before, a Gabidulin code $\calC \subseteq \Fqm^n$ with $\dr(\calC) = d$ defined by the parity-check matrix $H = \mat{h_j^{[i]}}$. Let $H' = \mat{\alpha^{[i+j]}}$, $0 \leq i \leq d-2$, $0 \leq j \leq n'-1$.
Then $H = H'A$ for some $A \in \Fq^{n' \times n}$ and some $n'$ satisfying $n \leq n' \leq m$. Thus, the map given by $c' = Ac$ is an injection from $\calC$ to $\calC'$, where $\calC' \in \Fqm^{n'}$ is the Gabidulin code defined by $H'$. Note that $\dr(\calC')=d$. Thus, a received word $r$ can be decoded by first applying a decoder for $\calC'$ on $r'=Ar$, yielding a codeword $c' = Ac$, and then computing $c = A^\dag c'$, where $A^\dag$ is a left inverse to $A$. The decoding complexity is equal to $2n' n m$ additions and multiplications in $\Fq$ plus the complexity of decoding $\calC'$. Thus, we will assume in the following that $h_i = \alpha^{[i]}$, $i=0,\ldots,n-1$. (Note that there is apparently no good reason for choosing a different $H$.)

Now, consider the syndrome computation in Step~\ref{step:syndromes}. We have
\begin{equation}\nonumber
  S_\ell = \sum_{i=0}^{n-1} r_i \alpha^{[i+\ell]}, \quad \ell=0,\ldots,d-2.
\end{equation}
It follows that the syndromes can be computed with only $(d-1)n(C(T)-m) + (d-1)(n-1)m \leq (d-1)nC(T)$ \emph{additions} in $\Fq$ (no multiplications).

Consider the computation of
\begin{equation}\nonumber
  \gamma_j = \Gamma(\alpha^{[j]}) = \sum_{i=0}^{\tau} \Gamma_i \alpha^{[i+j]}, \quad j=0,\ldots,m-1
\end{equation}
in Step~\ref{step:matrix-linear-map}. Similarly, this computation can be done simply with $\tau m (C(T)-m) + \tau m^2 = \tau m C(T)$ additions in $\Fq$.

Thus, the steps that were once the most demanding ones are now among the easiest to perform.

There are some additional savings. Note that $\ul{h}$ is now an identity matrix with $m-n$ additional all-zero columns at the right; thus the cost of computing $L_j$ from $X_j$ in Step~\ref{step:error-locations} reduces to zero. (In particular, if $n=m$, then $\ul{X_j} = L_j^T$, i.e., $L_j^T$ is precisely the vector representation of $X_j$ with respect to the normal basis.)

\begin{figure}
  \includegraphics[width=0.97\columnwidth]{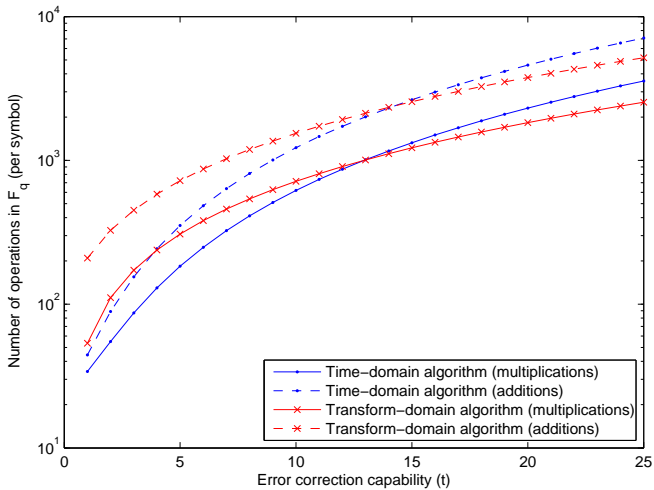}
  \caption{Complexity of the time-domain and transform-domain decoding algorithms, in operations per $\Fq$-symbol, as a function of the error correction capability $t$. An optimal self-dual normal basis is assumed. Parameters: $q = 256$, $n=m=51$, and $d=2t+1$.} \label{fig:comparison-complexity}
\vspace{-1ex}
\end{figure}

It follows that the decoding complexity is now dominated by Steps~\ref{step:ESP-ELP} and \ref{step:Gabidulin-algorithm} (although the kernel computation in Step~\ref{step:left-null-space} may become significant if $d$ is very small). For $n=m$ and $d = 2\tau+1$, the overall complexity of the algorithm is approximately $\frac{11}{2}\tau^2m^2 + \frac{1}{2}m^3$ multiplications and $\frac{11}{2}\tau^2mC(T)+\frac{1}{2}m^3$ additions in $\Fq$. An example is illustrated in Fig.~\ref{fig:comparison-complexity} for varying rates.

%% file: section4.tex
\section{Transform-Domain Methods}
\label{sec:transform-domain}

\subsection{Linear Maps over $\Fqm$ and the $q$-Transform}
\label{ssec:linear-maps-transform}

In this section, unless otherwise mentioned, all polynomials are $q$-polynomials over $\Fqm$ with $q$-degree smaller than $m$. If $v \in \Fqm^n$ is a vector of length $n \leq m$ over $\Fqm$, we will take $v$ to have length $m$, i.e., $v \in \Fqm^m$, and set $v_n=\cdots=v_{m-1}=0$.

We adopt the following convenient notation: if $f(x) = \sum_{i=0} f_i x^{[i]}$ is $q$-polynomial, then $f = \mat{f_0 & \cdots & f_{m-1}}^T$ is a vector over $\Fqm$, and vice-versa. Thus, $f(x)$ and $f$ are simply equivalent representations for the sequence $f_0,\ldots,f_{m-1}$. In addition, we adopt a cyclic indexing for any such a sequence: namely, we define $f_i = f_{i \bmod m}$ for all $i$. With this notation, we can write the symbolic multiplication $h(x) = f(x) \otimes g(x)$ as a cyclic ``$q$-convolution,'' namely,
  $h_\ell = \sum_{i=0}^{m-1} f_i g_{\ell-i}^{[i]}, \quad \ell=0,\ldots,m-1.$

We define the \emph{full $q$-reverse} of a $q$-polynomial $f(x)$ as the $q$-polynomial $\bar{f}(x) = \sum_{i=0}^{m-1} \bar{f}_i x^{[i]}$, where $\bar{f}_i = f_{-i}^{[i]}$, $\forall i$.



For the remainder of this subsection, $\calA = \{\alpha_i\}$, $\calB = \{\beta_i\}$ and $\Theta = \{\theta_i\}$ are bases for $\Fqm$ over $\Fq$, with \emph{dual bases} $\calA' = \{\alpha'_i\}$, $\calB' = \{\beta'_i\}$ and $\Theta' = \{\theta'_i\}$, respectively.
%
Recall that dual bases satisfy the property that $\Tr(\alpha_i \alpha'_j)$ is equal to 1 if $i=j$ and is equal to 0 otherwise, where $\Tr(x) = \sum_{\ell = 0}^{m-1} x^{[\ell]}$ is the trace function \cite{Lidl.Niederreiter}.

\medskip
\begin{lemma}
  $M = \mat{f(x)}_{\calA}^{\calB}$ $\;\iff\;$ $M^T = \mat{\bar{f}(x)}_{\calB'}^{\calA'}$.
\end{lemma}

\medskip
\begin{lemma}
   Suppose $\calA$ is a normal basis. Let $F \in \Fqm^m$ be such that $\mat{F}_\calB = \mat{f(x)}_{\calA}^\calB$. Then $f_i = F(\alpha'_i)$, $i=0,\ldots,{m-1}$. In particular, $\mat{f}_\Theta = \mat{F(x)}_{\calA'}^\Theta$.
\end{lemma}

\medskip
\begin{definition}
  The $q$-transform of a vector $f \in \Fqm^m$ (or a $q$-polynomial $f(x)$) with respect to a normal element $\alpha$ is the vector $F \in \Fqm^m$ (or the $q$-polynomial $F(x)$) given by $F_j = f(\alpha^{[j]}) = \sum_{i=0}^{m-1} f_i \alpha^{[i+j]}$, $j=0,\ldots,m-1$.
\end{definition}
\medskip

\begin{theorem}
  The inverse $q$-transform of a vector $F \in \Fqm^m$ (or a $q$-polynomial $F(x)$) with respect to $\alpha$ is given by $f_i = F({\alpha'}^{[i]}) = \sum_{j=0}^{m-1} F_j {\alpha'}^{[i+j]}$, $i=0,\ldots,m-1$. In other words, the inverse $q$-transform with respect to $\alpha$ is equal to the forward $q$-transform with respect to $\alpha'$.
\end{theorem}

\subsection{Implications to the Decoding of Gabidulin Codes}
\label{ssec:transform-decoding}

Recall the notations of Section~\ref{ssec:decoding-gabidulin}. Assume that $\calA = \{\alpha^{[i]}\}$ is a normal basis and that the Gabidulin code has parity-check matrix $H = \mat{\alpha^{[i+j]}}$.

As in the transform-domain decoding of Reed-Solomon codes, the equation $r = c + e$, or $r(x) = c(x) + e(x)$, is translated to the transform domain as $R(x) = C(x) + E(x)$, where $R(x)$, $C(x)$ and $E(x)$ are the $q$-transforms with respect to $\alpha$ of $r(x)$, $c(x)$ and $e(x)$, respectively. Now, the fact that $C_\ell = c(\alpha^{[\ell]}) = 0$, $\ell=0,\ldots,d-2$, implies that $S_\ell = R_\ell = E_\ell$, $\ell = 0,\ldots,d-2$. Note also that $\tilde{S}_\ell = \bar{E}_{\ell - d + 2}$, $\ell=0,\ldots,d-2$.

\medskip
\begin{lemma}\label{lem:coefficient-agreement}
  Let $\Gamma(x)$, $S(x)$ and $E(x)$ be linearized polynomials with $q$-degrees at most $\tau$, $d-2$ and $m-1$, respectively, and suppose and $E(x)$ agrees with $S(x)$ in the first $d-1$ coefficients. Then $\Gamma(x) \otimes E(x) \bmod x^{[m]}-x$ agrees with $\Gamma(x) \otimes S(x)$ in the coefficients $\tau \leq \ell \leq d-2$.
\end{lemma}
\medskip

\begin{theorem}[The Key Equations]
\begin{align}
\Gamma(x) \otimes E(x) &\equiv 0 \pmod{x^{[m]} - x} \nonumber \\
\Lambda(x) \otimes \bar{E}(x) &\equiv 0 \pmod{x^{[m]} - x}.\nonumber
\end{align}
In particular, (\ref{eq:key-equation-ESP}) and (\ref{eq:key-equation-ELP}) hold.
\end{theorem}
\begin{proof}
  For the first key equation, let $\ul{\gamma} = \mat{\Gamma(x)}_\calA^\calA$. Note that $\Gamma(V_j)=0$ implies $\ul{V_j}\, \ul{\gamma} =0$, $j=1,\ldots,\tau$. From (\ref{eq:matrix-multiplication-linear-maps}) we have
\begin{multline}\nonumber
  \mat{\Gamma(x) \otimes E(x)}_{\calA'}^\calA\\ = \mat{E(x)}_{\calA'}^\calA \mat{\Gamma(x)}_{\calA}^\calA = \ul{e}\, \ul{\gamma} = \sum_{j=1}^\tau \ul{X_j}^T\, \ul{V_j}\, \ul{\gamma} = 0.
\end{multline}
The form (\ref{eq:key-equation-ESP}) of this key equation follows immediately after applying Lemma~\ref{lem:coefficient-agreement}.

The proof of the second key equation is similar and is omitted due to lack of space.
%
\end{proof}
\medskip

Besides allowing us to give conceptually simpler proofs of the key equations, the transform approach also provides us with the theoretical ground for proposing a new decoding algorithm for Gabidulin codes. The main idea is that, after the ESP or the ELP is found, the remaining coefficients of $E(x)$ can be computed from
\begin{align}
  E_\ell = - \sum_{i=1}^{\tau} \Gamma_{i} E_{\ell-i}^{[i]} &= 0, \quad \ell = d-1,\ldots,m-1 \nonumber \\
  \bar{E}_\ell = - \sum_{i=0}^{\tau} \Lambda_{i} \bar{E}_{\ell-i}^{[i]} &= 0, \quad \ell = 1,\ldots,m-d+1. \nonumber
\end{align}
Then, the error polynomial $e(x)$ can be obtained through an inverse $q$-transform.

Computing this inverse transform takes, in general, $nm$ multiplications and additions in $\Fqm$ (or $km$ if the code is systematic and the parity portion is ignored). However, if $\calA$ is a self-dual normal basis, then an inverse transform becomes a forward transform, and the same computational savings described in Section~\ref{ssec:fast-decoding-normal-bases} can be obtained here. Note that most normal bases constructed via Gauss periods over fields of characteristic 2 are indeed self-dual (see Section~\ref{ssec:normal-bases} and, e.g., \cite{Gao++2000:AlgorithmsExponentiation}).

Below is a summary of the new algorithm, together with a breakdown of the complexity.

\begin{enumerate}
  \item \label{step:syndromes} \emph{Compute the syndromes:} see Section~\ref{ssec:decoding-gabidulin}.
  \item \label{step:ESP-ELP} \emph{Compute the ESP/ELP:} see Section~\ref{ssec:decoding-gabidulin}.
  \item \label{step:recursion} \emph{Compute $E(x)$ recursively:} $(m-d+1)\tau$ multiplications and $(m-d+1)(\tau-1)$ additions in $\Fqm$.
  \item \label{step:inverse-transform} \emph{Compute the error word:} $n m C(T)$ additions in $\Fq$ (or $k m C(T)$ if the code is systematic).
\end{enumerate}

%
%
%
%


As it can be seen from Step~\ref{step:recursion} above, the new algorithm essentially replaces the $O(d^2)$ operations of Gabidulin's algorithm with the $O(d(m-d))$ operations required for recursively computing $E(x)$. Thus, the algorithm is most beneficial for low-rate codes. For $n=m$ and $d=2\tau+1$, the overall complexity of the algorithm is approximately $(m+2\tau)\tau m^2$ multiplications and $(m+2\tau)\tau m C(T)$ additions in $\Fq$. It is straightforward to check that this complexity is smaller than that of \cite{Loidreau2005} (see \cite{Gadouleau.Yan2008a:Complexity}). An example is illustrated in Fig.~\ref{fig:comparison-complexity}. 

%% file: section5.tex
\section{Fast Encoding}
\label{sec:encoding}

As for any linear block code, encoding of Gabidulin codes requires, in general, $O(kn)$ operations in $\Fqm$, or $O(k(n-k))$ operations in $\Fqm$ if systematic encoding is used.

We show below that, if the code has a high rate and $\Fqm$ admits a low-complexity normal basis, then the encoding complexity can be significantly reduced. Alternatively, if nonsystematic encoding is allowed and $\Fqm$ admits a self-dual low-complexity normal basis, then very fast encoding is possible.

\subsection{Systematic Encoding of High-Rate Codes}
\label{ssec:systematic-encoding}

Let $c_{n-k},\ldots,c_{n-1} \in \Fqm$ denote the message coefficients. We set $r_i=0$, $i=0,\ldots,n-k-1$ and $r_i=c_i$, $i=n-k,\ldots,n-1$, and perform \emph{erasure decoding} on $r = \mat{r_0 & \cdots & r_{n-1}}^T$ to obtain $c_0,\ldots,c_{n-k-1}$.

We use the algorithm of Section~\ref{ssec:decoding-gabidulin}, with the computational savings of Section~\ref{ssec:fast-decoding-normal-bases}.
Note that only steps 1, 4 and 5 need to be performed, since the error locations (and thus also the error locators) are known: for $j=1,\ldots,d-1$, $L_j$ is a column vector with a 1 in the $j$th position and zero in all others. 
Thus, the complexity is dominated by Gabidulin's algorithm, requiring $O(d^2)$ operations in $\Fqm$ (see Section~\ref{ssec:decoding-gabidulin}). 
For high-rate codes, this improves on the previous value of $O(dk)$ mentioned above. Note that, without the approach in Section~\ref{ssec:fast-decoding-normal-bases}, encoding by erasure decoding would cost $O(dn)$ operations in $\Fqm$.

\subsection{Nonsystematic Encoding}
\label{ssec:nonsystematic-encoding}

Here we assume that $n=m$. Let $F_{m-k},\ldots,F_{m-1}$ denote the message coefficients, and let $F(x) = \sum_{j=m-k}^{m-1} F_j x^{[j]}$. We encode by taking the (inverse) $q$-transform with respect to $\alpha$, where $\calA = \{\alpha^{[i]}\}$ is a self-dual normal basis. Then $c_i = F(\alpha^{[i]})$, $i=0,\ldots,m-1$. It is clear that this task takes only $mkC(T)$ additions in $\Fq$, and is therefore extremely fast. The decoding task, however, has to be slightly updated.

Since, by construction, every codeword satisfies $c(\alpha^{[i]}) = 0$ for $i=0,\ldots,d-2$, most part of the decoding can remain the same. If decoding is performed in the time domain, then one additional step is needed to obtain the message: namely, computing the forward $q$-transform $F_j = c(\alpha^{[j]})$, for $j=m-k,\ldots,m-1$. These extra $mkC(T)$ additions in $\Fq$ barely affect the decoding complexity. On the other hand, if decoding is performed in the transform domain, than the last step (obtaining $e(x)$ from $E(x)$) can be simply skipped, as $F(x) = R(x) - E(x)$. This further saves at least $mkC(T)$ additions in $\Fq$.

%% file: section7.tex
\section{Conclusions}
\label{sec:conclusions}

In this paper, we have presented fast encoding and decoding algorithms for Gabidulin codes, both in time and in transform domain. The algorithms derive their speed from the use of an optimal (or low-complexity) normal basis, and the fact that multiplication by a $q$-power of $\alpha$ in such a normal basis can be performed very quickly. With respect to systematic high-rate codes (which seem to be the most suitable to practical applications), the decoding complexity is now dominated by the BM algorithm and Gabidulin's algorithm. An efficient implementation of these two algorithms is therefore an important practical question.